# m-sophistication

Bruno Bauwens [*]

Department of Electrical Energy, Systems and Automation,
Ghent University,
Technologiepark 913, B-9052, Zwijnaarde, Belgium,
Bruno.Bauwens@ugent.be.

**Abstract.** The $m$-sophistication of a finite binary string $x$ is introduced as a generalization of some parameter in the proof that complexity of complexity is rare. A probabilistic near sufficient statistic of $x$ is given which length is upper bounded by the $m$-sophistication of $x$ within small additive terms. This shows that $m$-sophistication is lower bounded by coarse sophistication and upper bounded by sophistication within small additive terms. It is also shown that $m$-sophistication and coarse sophistication can not be approximated by an upper or lower semicomputable function, not even within very large error.

## Introduction

The Kolmogorov complexity of a finite binary sequence is a measure for the amount of structure in a finite discrete sequence. Sophistication [1,17] is a measure to quantify the complexity of this structure. It is shown here that sophistication and its introduced variant $m$ sophistication is related to three important questions in the field of statistics and computability.

- If the Kolmogorov complexity $K(x)$ is low for some binary finite sequence $x$, than $x$ can be interpreted as "deterministically" generated, and "non-deterministically" generated otherwise. The structure function [16,19,22] defines for each $x$ a function of a natural number $k$ to the logarithm of the minimal cardinality of $x$ containing sets. If the structure function decreases for low $k$ to the value $K(x) - k$, these sequences are called "positively random". Positive randomness is satisfied with high probability if $x$ is "stochastically" generated. Such $x$ allow a useful definition of frequentistic probabilities satisfying the Kolmogorov probability axioms.

[*] Supported by a Ph.D grant of the Institute for the Promotion of Innovation through Science and Technology in Flanders (IWT-Vlaanderen). The work was carried out while the author was connected to the interdisciplinary Guislain research group at Ugent.

- A sumtest for a computable semimeasure is an abstraction of a statistical significance test for a simple hypothesis [19]. It can be argued that for many composite hypotheses, a theoretical ideal statistical test is given by a sumtest for a lower semicomputable semimeasures [4]. The question rises whether for the lower semicomputable semimeasure unbounded sumtests exists in some computability class. It turns out that for the hypotheses of independence there are no unbounded computable and lower semicomputable sumtests, but there are upper semicomputable sumtests of maximal magnitude $l(x)$ [7]. There are also no computable or lower semicomputable sumtests for a universal semimeasure, but there are upper semicomputable sumtests of magnitude $\log l(x) - O(\log \log l(x))$ [2]. The proof relies on the observation that the introduced $m$-sophistication for a universal semimeasure $m$, is within logarithmic terms a sumtest for $m$.
- The coding theorem justifies the approximation of the logarithm of a universal semimeasure by data-compression heuristics [10,11,21]. The hypothesis of a timeseries $x$ being influence-free of another timeseries $y$ corresponds to a universal online semimeasure [4,9]. Also the approximation of such a semimeasure is related to online complexities [4,9]. The error in such a coding result is given by $m$-sophistication [3,5].

*Overview and results.* The paper uses definitions and observations from [8] and basically runs through the proof of the theorem that high complexity of complexity is rare as in [13], see also [12,14,19]. $m$-sophistication is a generalization of a parameter used in this proof. It allows some simple observations related to the questions above. Let $k$ be the $m$-sophistication of a finite sequence $x$. It is shown that the amount $K(x)$ of information in $x$ can be decomposed as $k$ bits of Halting information and $K(x) - k$ bits of additional information, within $2 \log k$ error terms. The first $k$ bits of the Halting probability compute an approximate sufficient statistic for $x$. It is shown that within $O(\log k)$ terms $m$-sophistication is larger than coarse sophistication, and smaller than sophistication. Finally it is shown that $m$-sophistication and coarse sophistication define within logarithmic terms a sumtest relative to the universal semimeasure, and that they have no lower and upper semicomputable approximation, not even within large error.

*Definitions and notation.* For an introduction to Kolmogorov complexity and computability is refereed to [14,19] and for extensive specialized background to [12,20]. Let $\omega$ be the set of natural numbers. The binary strings $2^{<\omega}$ of finite length can be associated with $\omega$. Let $l(x)$ denote the length of $x$ in its binary expansion. Let $2^n$ and $2^{<n}$ be the sets of strings $x$ with $l(x) = n$, and $l(x) < n$. Let $\omega^{<\omega}$ be the set of finite sequences in $\omega$. The Real numbers in $[0,1]$ are associated with Cantor space[1]. For $r \in 2^\omega$, $r^k$ denotes $r_1 r_2 ... r_k$. For $x \in 2^{<\omega}$, $x^k$ denotes $x_1 x_2 ... x_k$.

A semimeasure $P$ is a positive Real function that satisfies $\sum \{P(x) : x \in \omega\} \leqslant 1$. A semimeasure $P$ (multiplicatively) dominates a semimeasure $Q$, notation:

---

[1] This association is not bijective since the Real $0.a0111...$ equals the Real $0.a1000...$ for any $a \in 2^{<\omega}$, however, this omission does not cause problems.

$P \geqslant^* Q$, if a constant $c$ exists such that for all $x$: $cP(x) \geqslant Q(x)$. $P =^* Q$, iff $P \leqslant^* Q$ and $Q \leqslant^* P$. A set $S$ of semimeasures has a universal element $m$ if $m \in S$ and $m$ dominates all semimeasures in $S$. Let $f, g$ be functions depending on parameters $x$ and $n$. $f$ dominates $g$ (notation: $f \geqslant^+ g$), iff there is a constant $c$ which satisfies for all $x$ and $n$: $f(x, n) + c \geqslant g(x, n)$. $c$ may depend on any parameter except $x, n$. $f =^+ g$ iff $f \leqslant^+ g$ and $g \leqslant^+ f$.

Let $\Phi(.|.)$ represent a fixed optimal universal Turing machine, that is prefix-free in its first argument. $\Phi_t(p|x) \downarrow = y$ means that $\Phi$ on input $p, x$ outputs $y$, and halts in less than $t$ computation steps. A Real function $f : \omega \to [0, 1]$ is computable if there is a $p \in 2^{<\omega}$ such that for all $k, x$: $\Phi(p|x, k) \downarrow = f(x)^k$. An enumeration of a Real function $f(x)$ is a computable real function $g(x, t)$ such that for all $t$: $g(u, t) \leqslant g(u, t + 1)$ and such that $\lim_{t,k} g(u, t) = f(u)$. A lower semicomputable function $f$ is a function that has an enumeration. A function $f$ is upper semicomputable if $-f$ is lower semicomputable. With abuse of notation, an enumeration of $f$ is denoted as $f_t$.

*Kolmogorov complexity and its properties.* For $x, y \in \omega^{<\omega}$, let the Kolmogorov complexity be

$$K_t(x) = \min\{l(p) : \Phi_t(p|y) \downarrow = x\}$$
$$K(x) = \lim_{t \to \infty} K_t(x).$$

For all $n \in \omega$: $K(n) \leqslant^+ \log n + 2 \log \log n$ and for all $x \in 2^n$: $K(x) \leqslant^+ n + 2 \log n$. Let $x^*$ represent the lexicographic first program that produces $x$.

$$K(x, y) =^+ K(y) + K(x|y^*) =^+ K(y) + K(x|y, K(y)).$$

A Halting program can also output its own length, therefore

$$K(x) =^+ K(x, K(x)).$$

The coding theorem shows that

$$Q_p(x) = \sum \{2^{-l(p)} : \Phi(p) \downarrow = x\} \tag{1}$$
$$Q_K(x) = 2^{-K(x)} \tag{2}$$

define universal semimeasures. This implies that for any universal semimeasure $m$: $-\log m(x) =^+ K(x)$.

## 1   Halting probability and a Buzzy Beaver variant

In computability theory, the number $\Omega$ is typically defined as the prior probability that some universal prefix-free Turing machine halts [8,13]. Here a closely related concept is studied: the probability that a universal semimeasure is defined.

**Definition 1.** *Let $m$ be some universal semimeasure.*

$$\Omega_{m,t} = \sum_{x < t} m_t(x)$$

$$\Omega_m = \lim_{t \to \infty} \Omega_t$$

The original definition in [8,13] is obtained by choosing $m = Q_p$, as in equation 1. $\Omega_{Q_p}$ satisfies the following well known theorem.

**Theorem 1.** *For all $n$: $K(\Omega_{Q_p}^n) \geqslant^+ n$. There is a constant $c$ such that for all $n$, the Halting of any program $p \in 2^{<n}$ can be decided by $\Omega^{n+c}$.*

It will be shown later in this section that these properties of $\Omega_{Q_p}$ remain for general $\Omega_m$ with a similar argument. Let $a, b$ represent objects or tuples of objects in $2^{<\omega}$ ($\omega$) that possibly depend on the parameters $n$ or $x$. It is said that "$a$ *computes* $b$" (notation: $a \longrightarrow b$) iff there is a constant $c$ that for all values of the parameters $x$ and $n$: $K(b|a) \leqslant c$. For $\alpha, \beta \in 2^\omega$, the relation $\alpha^n \longrightarrow \beta^n$ defines a partial order on $2^\omega$, which is equivalent with the 'domination' relation in [18]. $\Omega_{Q_p}$ is stable with respect to the choice of universal machine $\Phi$. Let $\Phi$ to $\Phi'$ be two optimal universal prefix-free Turing machines and let $Q_p$ and $Q_p'$ be defined as in equation 1, than it is easily observed that

$$\Omega_{Q_p}^n \longleftrightarrow \Omega_{Q_p'}^n.$$

An other example of such a relation is

$$\Omega_{Q_p}^n \longrightarrow \Omega_{Q_K}^n,$$

where $Q_K$ is defined in equation 2. It is an interesting question whether the opposite direction also holds.

Following the proof that high $K(K(x)|x)$ is rare in [14], the times $t_n$ are defined. Fix some universal semimeasure $m$, and let for each $n$:

$$t_n = \min\{t : \Omega_m - \Omega_{m,t} \leqslant 2^{-n}\}.$$

It is easily observed that

**Lemma 1.**

$$\Omega_m^n \longleftrightarrow n, t_n$$

**Lemma 2.** *Let $t[p] = \min_t\{\phi_t(p) \downarrow\}$. For any universal $m$, there is a constant $c$ such that for any halting $p \in 2^{<\omega}$:*

$$\Phi(p) \leqslant t_{l(p)+c}$$

$$t[p] \leqslant t_{l(p)+c}.$$

*Proof.* Let $n = l(p) + c + 1$ with $c$ large enough and suppose that $l(p) \geqslant c + 2$. Let $x \in 2^{<\omega}$ be the lexicographic first string with $-\log m_{t_n}(x) \geqslant l(x) \geqslant 2l(p)$. Suppose that $\phi(p) \geqslant t_n$, than $p \longrightarrow p, n \longrightarrow x$ and thus

$$-\log m(x) \leqslant^+ K(x) \leqslant^+ l(p).$$

which implies for $c$ sufficiently large

$$\Omega - \Omega_{t_n} \geqslant m(x) - m_{t_n}(x) \geqslant 2^{-l(p)-c} - 2^{-2l(p)} > 2^{-l(p)-c-1},$$

contradicting the definition of $t_n$. The second claim follows by remarking that for every Halting $p$: $p \longrightarrow t[p]$. □

The Buzzy Beaver function is defined by:

$$BB(n) = \max\{\Phi(p) : l(p) \leqslant n\}.$$

Lemma 3 shows that $t_n$ is a very fast growing function that oscillates between $BB(n)$ and $BB(n + 2\log n)^2$.

**Lemma 3.** *There exists a constant $c$ such that:*

$$BB(n - c) \leqslant t_n < BB(n + 2\log n + c).$$

*Proof.* The left inequality follows from Lemma 2. By Lemma 1

$$K(t_n) \leqslant^+ K(\Omega_m^n) \leqslant^+ n + K(n),$$

The witness of $K(t_n)$ shows the right inequality. □

**Corollary 1.** *For all universal semimeasures $m$, $m'$ there is some constant $c$ such that*

$$t_n < t'_{n+2\log n+c},$$

*with $t_n$ and $t'_n$ defined by $m$ and $m'$.*

*Proof.*

$$t_n \leqslant BB(n + 2\log n + c) < t'_{n+2\log n+2c}$$

□

A Real number $\alpha \in 2^\omega$ is random if for any $n$: $K(\alpha^n) \geqslant^+ n$. It follows by Lemma 3 that

**Corollary 2.** *$\Omega_m$ is random.*

*Proof.* Since $n \leqslant^+ K(t_n) \leqslant^+ K(\Omega_m^n)$. □

By Corollary 1 it follows that

---

[2]  Remark that analogue bounds as in Lemma 5 can be proved.

**Lemma 4.** *for $m, m'$ universal semimeasures*

$$\Omega_m^n \longrightarrow \Omega_{m'}^{n-2\log n}.$$

*Proof.*

$$\Omega_m^n \longrightarrow n, t_n \longrightarrow n, t'_{n-2\log n} \longrightarrow \Omega_{m'}^{n-2\log n}$$

□

The question rises whether the set of all $\Omega_m$ for some universal semimeasures has a maximal element relative to the $\longrightarrow$ order. Remark that it is shown in [18] that the set of all $\Omega_m$ with $m$ universal corresponds to all computable enumerable random Real numbers.

Finally it can be asked whether these logarithmic bounds are tight. Some remarks are made in relation to this question. For a random $\alpha \in 2^\omega$ only a small amount of values $K(\alpha^n)$ is allowed:

$$n \leqslant^+ K(\alpha^n) \leqslant^+ n + 2\log n.$$

It is well known that $K(\alpha^n)$ oscillates within these bounds.

**Lemma 5.** *For any random $\alpha \in 2^\omega$ there are an infinite amount of $n$ such that*

$$K(\alpha^n) \leqslant^+ n + 2\log\log n,$$

*and there are an infinite amount of $n$ such that*

$$K(\alpha^n) \geqslant^+ n + \log n.$$

See appendix for the proof.

## 2   $m$-sophistication and complexity of complexity

**Definition 2.** *For some universal semimeasure $m$, and some $c \in \omega$, the $m$-sophistication an $x \in 2^{<\omega}$ is given by:*

$$k_c(x) = \min\{k : K_{t_k}(x) \leqslant K(x) + c\}.$$

$k_c(x)$ is limit-computable in $x$, but not lower semicomputable or upper semicomputable by Proposition 1. From Corollary 1 it is observed that $k_c$ is relatively stable with respect to changes of universal semimeasure $m$.

**Corollary 3.** *Let $m, m'$ be universal semimeasures and let $k$ and $k'$ be the $m$-sophistication and $m'$-sophistication, then for any $c$:*

$$k_c \leqslant^+ k'_c + 2\log k'_c.$$

As for sophistication (see further), also $m$-sophistication is unstable with respect to the parameter $c$.

**Lemma 6.** *There is a $c'$ such that for all $c$ there are infinitely many $x$ with*

$$k_c(x) - k_{c+c'}(x) \geqslant^+ n - 2\log n.$$

See appendix for the proof.

By the coding theorem, a definition very related to $m$-sophistication is given by $(m,m)$-sophistication:

$$k'(x) = \min\{k : \frac{m(x)}{m_{t_k}(x)} \leqslant 2\}.$$

**Lemma 7.** *For any $c$ large enough: $k' \geqslant k_c$.*

*Proof.* By some time-bounded version of the coding theorem:

$$K_{t_{k'(x)+c}}(x) \leqslant^+ -\log m_{t_{k'(x)}}(x) =^+ -\log m(x) =^+ K(x).$$

$\square$

High $(m,m)$-depth is rare.

**Lemma 8.** *For any $k$ and $S_k = \{x : k'(x) \geqslant k\}$:*

$$m(S_k) \leqslant 2^{-k+1}.$$

*Proof.*

$$\frac{1}{2}m(S_k) \leqslant m(S_k) - m_{t_k}(S_k) \leqslant \Omega - \Omega_{t_k} \leqslant 2^{-k}.$$

$\square$

**Lemma 9.** *Let $k(x)$ be either $k'(x)$ or $k_c(x)$ for any $c$, than:*

$$K(K(x)|x) \leqslant^+ k(x) + 2\log k(x).$$

*Proof.* Remark that $t_{k(x)}, x \longrightarrow K(x)$, thus

$$K(K(x)|x) \leqslant^+ K(t_{k(x)}) \leqslant^+ K(\Omega^{k(x)}) \leqslant^+ k(x) + 2\log k(x),$$

where the last inequality follows from Lemmas 1 and 5. $\square$

**Corollary 4.** *There exists a constant $c > 0$ such that*

$$m(\{K(K(x)|x) \geqslant k\}) \leqslant c2^{-k-2\log k}.$$

A sumtest $d$ for a semimeasure $P$ is a function $d : 2^{<\omega} \to \mathbb{Z}$ such that

$$\sum_{x \in \omega} P(x) 2^{d(x)} \leqslant 1.$$

**Corollary 5.** *For $k = k'$ and for $k = k_c$ with $c$ large enough, $k - 2\log k$ defines a sumtest for $m$.*

*Proof.*

$$\sum_{x \in 2^{<\omega}} m(x) 2^{k'(x) - 2\log k'(x) - 2} \leqslant \sum_{k \in \omega} m(S_k) 2^{k - 2\log k - 2} \leqslant \sum_{k \in \omega} 2^{-2\log k - 1} \leqslant 1$$

$\square$

$k_c$ and $k'$ are not computable, and not even a logarithmic lower bound can be computed.

**Proposition 1.** *For $k = k'$ and for $k = k_c$ with $c$ large enough, $k$ can not be approximated by a lower or upper semicomputable function within $k - 2\log k + O(1)$ error.*

See appendix for the proof.

## 3   Sophistication and coarse sophistication

Let $f$ be a computable function. A function $f$-*sufficient statistic* [15] is a computable prefix-free function $g$ such that there exists a $d \in g^{-1}(x)$ with

$$K(g) + l(d) \leqslant K(x) + f(l(x)).$$

The sophistication [17] of $x \in 2^{<\omega}$ is given by:

$$k_c^{\mathrm{soph}}(x) = \min\{K(f) : f \text{ is a } c\text{-sufficient statistic of } x\}.$$

Remark that there is a slight deviation from [17,23] since it is also required that $f$ is prefix-free. This is necessary to interpret sophistication as the length of a minimal sufficient statistic [15]. Also remark that now Lemma 10 is true. Let $bb(x)$ be the inverse of the Buzzy Beaver function, it is $bb(x) = \min\{k : x \leqslant BB(k)\}$. It is a very slow growing function, dominated by any unbounded non-decreasing function [7].

**Proposition 2.** *There exists a $c'$ such that for all $c, x$:*

$$k_{c+c'}(x) \leqslant^+ k_c^{\mathrm{soph}}(x) + bb(x).$$

*Proof.* The right inequality follows by observing that any function $f$, witnessing the definition of sophistication defines a description of $x$ of length $K(x) + c + c'$, for some $c'$ large enough. Let $d = \min\{d : f(x) = d\}$, let

$$M = BB(bb(x)) \geqslant x \geqslant d,$$

and let $p$ be the program that evaluates $f(e)$ for all $e \leqslant M$. Let $s$ be the computation time of this computation. Remark that $K_s(x) \leqslant K(x) + c + c'$ and thus

$$s \geqslant t_{k_{c+c'}(x)} \geqslant BB(k_{c+c'} - c')$$

for some $c'$ large enough, by Lemma 3. This implies

$$k_{c+c'} \leqslant^+ K(s) \leqslant^+ l(p) \leqslant^+ K(f) + bb(x) \leqslant^+ k^{\mathrm{soph}}(x) + bb(x).$$

$\square$

A probabilistic $f$-sufficient statistic is a computable probability distribution[3] $P$ such that

$$K(P) - \log P(x) \leqslant K(x) + f(l(x)).$$

Since prefix-free functions are used here, probabilistic and function sufficient statistics are equivalent.

**Lemma 10.** *There is a constant $c$ such that every probabilistic $f$-sufficient statistic $P$ defines a function $(f + c)$-sufficient statistic $g$ with $abs(K(P) - K(g)) \leqslant c$, and every function $f$-sufficient statistic $g$ defines a probabilistic $(f + c)$-sufficient statistic $P$ with $abs(K(P) - K(g)) \leqslant c$.*

*Proof.* The first claim is proved in [23]. It remains to show the second claim. Let $g$ be the function $f$-sufficient statistic and let

$$P(x) = \sum \{2^{-l(d)} : g(d) = x \wedge d \leqslant x\}.$$

Remark that $P(x) = 0$ if there is no $d \leqslant x$ with $g(d) = x$. It follows that $-\log P(x) \leqslant l(d)$ for the witness $d$ of $x$ in the definition of the function $f$-sufficient statistic of $g$. Remark that $K(g) \leqslant^+ K(P)$, and therefore the conditions of the definition of $(f + c)$-sufficient statistic are fulfilled.     $\square$

Let

$$P_k(x) = N2^{-k}(m_{t_k}(x) - m_{t_{k-1}}(x)),$$

Where $N$ is a normalization constant such that $P_k$ defines a computable probability distribution. Remark that $2 \leqslant N < 4$. Also remark that this can be considered as the probabilistic equivalent of the "explicit minimal near sufficient set statistic" described in [15].

**Lemma 11.** *For $m = Q_K$:*

$$K(x|\Omega^{k'(x)}) \leqslant^+ K(x) - k'(x).$$

*Proof.* Remark that since $m = Q_K$, for any $k$ either $m_{t_k}(x) = m_{t_{k-1}}(x)$ or $m_{t_k}(x) = 2m_{t_{k-1}}(x)$. This implies that $P_{k'(x)}(x) = 2^{-K(x)-1}$. The Lemma follows by Shannon-Fano coding.     $\square$

To relate $P_k$ to sophistication, it is shown that it defines some $f$-sufficient statistic.

**Proposition 3.** *There exists a $c$ such that $P_{k'(x)}$ is a probabilistic $(2 \log k'(x) + c)$-sufficient statistic for $x$. There exists a $c$ such that for any $c'$, there is a $k \leqslant^+ k_{c'}(x)$ such that $P_k$ is a $(3 \log k_c(x) + c + c')$-sufficient statistic for $x$.*

---

[3]  A *probability distribution* is a semimeasure with $\sum_{x \in \omega} P(x) = 1$

See appendix for the proof.

The online coding theorem [9] relates the logarithm of a universal online semimeasures (causal semimeasure) to online Kolmogorov complexity. The online coding theorem has an error term, which is improved for the length-conditional case in [3,5]. In the proof of the improved online coding theorem, an online computable semimeasure is associated with $P_{k'(x)}$. It is shown that the value of the logarithm of the universal online semimeasure and the associated semimeasure for $x$ equals within a $O(\log k'(x))$ term. Since the associated semimeasure is computable, a variant of Shannon-Fano code can be applied.

In [6] it is shown that the result of Proposition can not be further improved to eliminate the logarithmic terms in order to consider $P_k$ as a probabilistic $c$-sufficient statistic. It is shown that minimal sufficient statistics contain a substantial amount of non-Halting information. The proof seems to imply that in contrast with $m$-sophistication, sophistication does not define a sumtest. However, it is shown in [6] that $P_k$ defines a minimal typical model [24].

Sophistication is unstable with respect to the parameter $c$, therefore in [1] coarse sophistication is defined as

$$k^{\mathrm{csoph}}(x) = \min_c \{ k_c(x) + c \}.$$

As a corollary of Proposition 3 it follows that:

**Corollary 6.**
$$k^{\mathrm{csoph}}(x) \leqslant^+ k'(x) + 2\log k'(x).$$

**Proposition 4.** $k^{\mathrm{csoph}}(x) - 4\log k^{\mathrm{csoph}}(x)$ *defines a sumtest for* $m$. $k^{\mathrm{csoph}}$ *can not be approximated by a lower or upper semicomputable function within* $k - 2\log k + O(1)$ *error.*

*Proof.* This follows from Corollary 6 and the same proof as 1. ☐

**Acknowledgments** The author is also grateful to M.Li and P.Vitanyi, for the book [19], without such a good introductory and reference book this work would never have appeared.

## Appendix: Proofs of some lemmas and propositions

*Proof of Lemma 5.* Let for any $k$ $n = 1^\frown \alpha^k$ such that $n$ can be computed by $\alpha^k$ and such that $\log n =^+ k$. Remark that for any $z \in 2^{n-k}$ one has

$$K(z|\alpha^k) \leqslant^+ K(z|n-k) \leqslant^+ n-k,$$

and consequently,

$$K(\alpha^n) \leqslant^+ K(\alpha^k) + K(\alpha_{k\ldots n}|\alpha^k)$$
$$\leqslant^+ k + 2\log k + n - k.$$

The second inequality follows from Exercise 3.6.3d in [19].    □

*Proof of Lemma 6.* Kolmogorov complexity fluctuates "continiuously", in the sense that there exists a constant $e$ such that for all $a, r$ $K(r+1, a) - e \leqslant K(r, a) \leqslant K(r+1, a)$. Let $e$ be such a constant large enough. Since $K(t_{n-2\log n - c - 2e}) \leqslant n - c - e$, there always exists an $r$ such that:

$$n - c - 2e \leqslant K(r, t_{n-2\log n - c - 2e}) < n - c - e.$$

Remark that $r \leqslant n^3$ can be chosen for $n$ large enough. Let $t = t_{n-2\log n - c - 2e}$ and let $x \in 2^n$ the lexicographic $r$-th string such that $K_t(x) =^+ n$. Remark that such an $x$ always exists by Lemma 12 and

$$t, r, n \longleftrightarrow x.$$

This implies that for $e$ large enough:

$$n - c - 3e \leqslant K(x) < n - c.$$

Therefore

$$c < K_t(x) - K(x) \leqslant c + 3e.$$

    □

**Lemma 12.** *For some computable function $f$ large enough, and some constant $c$ large enough, there are infinitely many $n$, such that the amount of $x \in 2^n$ with $n - c \leqslant K_{f(n)}(x) \leqslant n + c$ is larger than $2^{n-2\log n}$.*

*Proof.* There are infinitely many $m$ such that $K_{f(m)}(m) =^+ K(m)$. For any such $m$ let $n = K(m) + m$. Remark that that there are $2^m$ many $r \in 2^m$ such that $K(r|m^*) =^+ r$, with $m^*$ a shortest program for $m$. Let $r' \in 2^n$ be $r \in 2^{n-K(m)}$ preappended with $m^*$. This shows that $r' \longleftrightarrow m, r$ and thus

$$n = K(m) + m =^+ K(m) + K(r|m^*) =^+ K(r').$$

Thus $K_{f^{(2)}(n)}(r') \geqslant^+ n$. Also

$$K_{f^{(2)}(n)}(r') \leqslant^+ K_{f(m)}(m) + K_{f(m)}(r|m^*) =^+ n,$$

for $f$ large enough.    □

*Proof of proposition 1.* Suppose that the function $d$ approximates $k$ such that $k - d \leqslant k - e \log k + O(1)$ for some constant $e$. This implies that $d \geqslant e \log k - O(1)$. Remark that this implies by Corollary 5 that there exists a $c'$ such that $d - 4 \log d - c'$ is a sumtest for $m$.

By [7] every lower semicomputable sumtest for $m$ is bounded by a constant, which implies that if $d$ was lower semicomputable, than $d \leqslant^+ 0$, and thus only the constant $e = 0$ is allowed.

By [2] every upper semicomputable sumtest for $m$ is bounded by a $\log l(x) + O(\log \log l(x))$. Therefore, only the constant $e = 1$ is allowed. □

*Proof of Proposition 3.* Remark that for any $k$: $K(P_k) \leqslant^+ k + 2 \log k$. Choosing $k = k'(x)$, and remarking that $-\log P_{k'(x)}(x) =^+ K(x) - k'(x)$, proves the first claim.

The second claim is now proved. By some time bounded version of the coding theorem there is a constant $e$ such that:

$$\log m(x) =^+ K(x) \leqslant K_{t_{k_c}}(x) + c \leqslant^+ -\log m_{t_{k_c(x)+e}}(x).$$

Therefore

$$m(x) \leqslant^* m_{t_{k_c(x)+e}}(x) =^* 2^k \sum \{ P_k(x) : k \leqslant k_c + e \}.$$

This shows that there is a $k$ such that

$$m(x) \leqslant^* \frac{2^k}{k} P_k(x).$$

By applying the coding theorem, and taking $-\log$ of the above equation one obtains:

$$\begin{aligned}
K(x) &=^+ -\log m(x) \\
&\geqslant^+ k - \log k - \log P_k(x) \\
&\geqslant^+ K(P) - 3 \log k - \log P_k(x).
\end{aligned}$$

Which shows that $P_k$ is a $(3 \log k + e')$-sufficient statistic. Remark that $e' \leqslant c + c'$ for some $c'$ independent of $c$. □